\documentclass[12pt]{article}
\usepackage{amsmath,amssymb}
\usepackage{graphicx}
\usepackage[pdftex]{hyperref}
\DeclareGraphicsExtensions{.eps,.bmp,.wmf,.jpeg,.pdf}
\numberwithin{equation}{section}
\def\be{\begin{equation}}
\def\ee{\end{equation}}

\def\bea{\begin{eqnarray}}
\def\eea{\end{eqnarray}}

\title{General Non-minimal Kinetic coupling to gravity}
\author{L.N. Granda\thanks{ngranda@univalle.edu.co} and W. Cardona\thanks{wilalbca@univalle.edu.co}\\ {\small\it Departamento de Fisica, Universidad del Valle}} 
\date{}
\begin{document}
\maketitle

\begin{abstract}
\noindent We study a new model of scalar field with a general non-minimal kinetic coupling to itself and to the curvature, as a source of dark energy, and analyze the cosmological dynamics of this model and the issue of accelerated expansion. A wide variety of scalar fields and potentials giving rise to power-law expansion have been found. The dynamical equation of state is studied for the two cases, without and with free kinetic term . In the first case, a behavior very close to that of the cosmological constant was found. In the second case, a solution was found, which match the current phenomenology of the dark energy. The model shows a rich variety of dynamical scenarios.\\ 

\end{abstract}

\section{Introduction}
\noindent 
The astrophysical data from distant Ia supernovae observations \cite{riess}, \cite{perlmutter}, \cite{kowalski}, cosmic microwave background anisotropy \cite{spergel1}, and large scale galaxy surveys \cite{tegmark},  \cite{abazajian1}, \cite{tegmark2}, all indicate that the current Universe is not only expanding, it is accelerating due to some kind of  negative-pressure form of matter known as dark energy (\cite{copeland}, \cite{sahnii}, \cite{padmanabhan}). The combined analysis of cosmological observations also suggests that the universe is spatially flat, and  consists of about $\sim 1/3$ of dark matter, and $\sim 2/3$ of homogeneously distributed dark energy with negative pressure. This dark energy may consist of cosmological constant, conventionally associated with the energy of the vacuum  or alternatively, could came from a dynamical varying scalar field at late times which also account for the missing energy density in the universe. That kind of scalar fields are allowed from several theories in particle physics and in multidimensional gravity, like Kaluza-Klein theory, String Theory\cite{green} or Supergravity, in which the scalar field appears in a natural way.
The cosmological solutions using scalar fields have been a subject of intensive study in the last years in the usual model called quintessence \cite{RP}, \cite{wett}, \cite{shani}, \cite{stein}, \cite{copeland97}, \cite{stein1}, \cite{stein2}, where the stability of scaling solutions was studied (i.e. solutions where the scalar energy density scale in the same way like a barotropic fluid) for a field evolving in accordance with an exponential potential and a power law potential, which provide a late time inflation. More exotic approaches to the problem of dark energy using scalar fields are related with K-essence models, based on scalar field with non-standard kinetic term \cite{stein3},\cite{chiba}; string theory fundamental scalars known as tachyons \cite{pad}; scalar field with negative kinetic energy, which provides a solution known as phantom dark energy \cite{caldwell} (see \cite{copeland} for a review). The non-minimal coupling between the quintessence field and curvature, as an explanation of the accelerated expansion, has been considered among others, in refs. \cite{chiba1} \cite{uzan},\cite{wands}. This coupling has also been studied in the context of the inflationary cosmology \cite{amendola1}, \cite{maeda},  \cite{turner}, \cite{kasper}, \cite{easson1}. An inflationary model including the DBI term and non-minimally coupled scalar has been proposed in \cite{easson2,easson3}.\\
\noindent Considering the possibility of having a situation where the accelerated expansion
arises out of modifications to the kinetic energy of the scalar field, in this paper we consider an explicit coupling between the scalar field, the kinetic term and the curvature, as a source of dark energy, and analyze the role of this new coupling in an evolution scenario with late-time accelerated expansion. The basic motivation for studying such theories is related with the fact that they appear as low energy limit of several higher dimensional theories, e.g. superstring theory \cite{green}, and provide a possible approach to quantum gravity from a perturbative point of view. They appear as part of the Weyl anomaly in $N=4$ conformal supergravity \cite{tseytlin, odintsov2}. A model with non-minimal derivative couplings was proposed in \cite{amendola2}, \cite{capozziello1}, \cite{capozziello2} in the context of inflationary cosmology, and recently, non-minimal derivative
coupling of the Higgs field was considered in \cite{germani}, also as inflationary model. In \cite{caldwell1} a derivative coupling to Ricci tensor has been considered to study cosmological restrictions on the coupling parameter, and the role of this coupling during inflation. Some asymptotical solutions for a non-minimal kinetic coupling to scalar and Ricci curvatures were found in \cite{sushkov}, and quintessence and phantom cosmological scenarios with non-minimal derivative coupling have been studied in \cite{saridakis}. 
Non-minimal coupling of arbitrary function $f(R)$ with matter Lagrangian (including the kinetic scalar term) has been introduced in \cite{sergei},\cite{allemandi}. Such a model was proposed to describe the dark energy, late-time universe acceleration. When $f(R)$ represents the power law function, such a model (which maybe considered as string-inspired theory) maybe proposed for dynamical resolution of cosmological constant problem as it has been suggested in \cite{sergei1},\cite{sergei2}. In the present work we consider this function $f(R)$ as linear in $R$ but we 
generalize the model permitting extra $R_{\mu\nu}$ coupling with kinetic-like scalar term. In addition, we keep the free kinetic term for the scalar field.\\
\noindent We propose a model of dark energy in which the scalar-field kinetic term is non-minimally coupled to itself and to the scalar and Ricci curvatures. We consider a general self coupling $F(\phi)$, which allows to find a wide variety of potentials giving rise to accelerated expansion. Some solutions for the product $F(\phi)\dot{\phi}^2$ are proposed, and the scalar field and potential are found. We have considered power-law solutions, and solutions giving rise to dynamically varying equation of state. In the last section we present some conclusions.

\section{Field Equations}

Let us start with the following  action, which generalizes the model proposed in \cite{granda}:

\be\label{eq1}
\begin{aligned}
S=&\int d^{4}x\sqrt{-g}\Big[\frac{1}{16\pi G} R-\frac{1}{2}\partial_{\mu}\phi\partial^{\mu}\phi-\frac{1}{2} \xi R \left(F(\phi)\partial_{\mu}\phi\partial^{\mu}\phi\right) -\\ 
&\frac{1}{2} \eta R_{\mu\nu}\left(F(\phi)\partial^{\mu}\phi\partial^{\nu}\phi\right) - V(\phi)\Big] + S_m.
\end{aligned}
\ee

\noindent where $S_m$ is the dark matter action which describes a fluid with barotropic equation of state. The dimensionality of the coupling constants $\xi$ and $\eta$ depends on the type of function $F(\phi)$. Taking the variation of action \ref{eq1} with respect to the metric, we obtain a general expression of the form 
\be\label{eq2}
R_{\mu\nu}-\frac{1}{2}g_{\mu\nu}R=\kappa^2\left[T_{\mu\nu}^m+T_{\mu\nu}\right]
\ee
where $\kappa^2=8\pi G$, $T_{\mu\nu}^m$ is the usual energy-momentum tensor for matter component, and the tensor $T_{\mu\nu}$ represents the variation of the terms which depend on the scalar field $\phi$ and can be written as
\be\label{eq3}
T_{\mu\nu}=T_{\mu\nu}^{\phi}+T_{\mu\nu}^{\xi}+T_{\mu\nu}^{\eta}
\ee
where $T_{\mu\nu}^{\phi}$, $T_{\mu\nu}^{\xi}$, $T_{\mu\nu}^{\eta}$ correspond to the variations of the minimally coupled terms, the $\xi$ and the $\eta$ couplings respectively. Due to the interacting terms this expressions are defined in the Jordan frame and do not correspond to the energy-momentum tensors as defined in the Einstein frame. Those variations are given by
\be\label{eq4}
T_{\mu\nu}^{\phi}=\nabla_{\mu}\phi\nabla_{\nu}\phi-\frac{1}{2}g_{\mu\nu}\nabla_{\lambda}\phi\nabla^{\lambda}\phi
-g_{\mu\nu}V(\phi)
\ee
\be\label{eq5}
\begin{aligned}
T_{\mu\nu}^{\xi}=&\xi\Big[\left(R_{\mu\nu}-\frac{1}{2}g_{\mu\nu}R\right)\left(F(\phi)\nabla_{\lambda}\phi\nabla^{\lambda}\phi\right)+g_{\mu\nu}\nabla_{\lambda}\nabla^{\lambda}\left(F(\phi)\nabla_{\gamma}\phi\nabla^{\gamma}\phi\right)\\
&-\frac{1}{2}(\nabla_{\mu}\nabla_{\nu}+\nabla_{\nu}\nabla_{\mu})\left(F(\phi)\nabla_{\lambda}\phi\nabla^{\lambda}\phi\right)+R\left(F(\phi)\nabla_{\mu}\phi\nabla_{\nu}\phi\right)\Big]
\end{aligned}
\ee
\be\label{eq6}
\begin{aligned}
T_{\mu\nu}^{\eta}=&\eta\Big[F(\phi)\left(R_{\mu\lambda}\nabla^{\lambda}\phi\nabla_{\nu}\phi+R_{\nu\lambda}\nabla^{\lambda}\phi\nabla_{\mu}\phi\right)-\frac{1}{2}g_{\mu\nu}R_{\lambda\gamma}\left(F(\phi)\nabla^{\lambda}\phi\nabla^{\gamma}\phi\right)\\
&-\frac{1}{2}\left(\nabla_{\lambda}\nabla_{\mu}\left(F(\phi)\nabla^{\lambda}\phi\nabla_{\nu}\phi\right)+\nabla_{\lambda}\nabla_{\nu}\left(F(\phi)\nabla^{\lambda}\phi\nabla_{\mu}\phi\right)\right)\\
&+\frac{1}{2}\nabla_{\lambda}\nabla^{\lambda}\left(F(\phi)\nabla_{\mu}\phi\nabla_{\nu}\phi\right)+\frac{1}{2}g_{\mu\nu}\nabla_{\lambda}\nabla_{\gamma}\left(F(\phi)\nabla^{\lambda}\phi\nabla^{\gamma}\phi\right)\Big]
\end{aligned}
\ee
Variating with respect to the scalar field gives rise to the equation of motion
\be\label{eq7}
\begin{aligned}
&-\frac{1}{\sqrt{-g}}\partial_{\mu}\left[\sqrt{-g}\left(\xi R F(\phi)\partial^{\mu}\phi+\eta R^{\mu\nu}F(\phi)\partial_{\nu}\phi+\partial^{\mu}\phi\right)\right]+\frac{dV}{d\phi}+\\
&\frac{dF}{d\phi}\left(\xi R\partial_{\mu}\phi\partial^{\mu}\phi+\eta R_{\mu\nu}\partial^{\mu}\phi\partial^{\nu}\phi\right)=0
\end{aligned}
\ee
Assuming the spatially-flat Friedmann–Robertson–Walker (FRW) metric,
\be\label{eq8}
ds^2=-dt^2+a(t)^2\left(dr^2+r^2d\Omega^2\right)
\ee
From Eqs. (\ref{eq4}-\ref{eq6}) and using the metric (\ref{eq8}) we can write the $(00)$ and $(11)$ components of the Eq. (\ref{eq2})  (with the Hubble parameter $H$, and for homogeneous time-depending scalar field) as follows
\be\label{eq9}
\begin{aligned}
H^2=&\frac{\kappa^2}{3}\Big[\frac{1}{2}\dot{\phi}^2+V(\phi)+9\xi H^2F(\phi)\dot{\phi}^2+3(2\xi+\eta)\dot{H}F(\phi)\dot{\phi}^2\\
&-3(2\xi+\eta)H F(\phi)\dot{\phi}\ddot{\phi}-\frac{3}{2}(2\xi+\eta)H \frac{dF}{d\phi}\dot{\phi}^3\Big]
\end{aligned}
\ee
and
\be\label{eq10}
\begin{aligned}
-2\dot{H}-3H^2=&\kappa^2\Big[\frac{1}{2}\dot{\phi}^2-V(\phi)+3(\xi+\eta)H^2F(\phi)\dot{\phi}^2+2(\xi+\eta)\dot{H}F(\phi)\dot{\phi}^2\\
&+4(\xi+\eta)H F(\phi)\dot{\phi}\ddot{\phi}+2(\xi+\eta)H\frac{dF}{d\phi}\dot{\phi}^3\\
&+(2\xi+\eta)\left(F(\phi)\ddot{\phi}^2+F(\phi)\dot{\phi}\dddot{\phi}+\frac{5}{2}\frac{dF}{d\phi}\dot{\phi}^2\ddot{\phi}+\frac{1}{2}\frac{d^2F}{d\phi^2}\dot{\phi}^4\right)\Big]
\end{aligned}
\ee
where we have assumed scalar field dominance (i.e. $T^m_{\mu\nu}=0$). The equation of motion for the scalar field (\ref{eq7}) takes the form
\be\label{eq11}
\begin{aligned}
&\ddot{\phi}+3H\dot{\phi}+\frac{dV}{d\phi}+3(2\xi+\eta)\ddot{H}F(\phi)\dot{\phi}+
3(14\xi+5\eta)H\dot{H}F(\phi)\dot{\phi}\\
&+\frac{3}{2}(2\xi+\eta)\dot{H}\left(2F(\phi)\ddot{\phi}+\frac{dF}{d\phi}\dot{\phi}^2\right)+
\frac{3}{2}(4\xi+\eta)H^2\left(2F(\phi)\ddot{\phi}+\frac{dF}{d\phi}\dot{\phi}^2\right)\\
&+9(4\xi+\eta)H^3F(\phi)\dot{\phi}=0
\end{aligned}
\ee
where the first three terms correspond to the minimally coupled field. In what follows we study the cosmological consecuences of this equations, under some conditions that simplify the search for solutions, but nevertheless show the richness of the cosmological dynamics of the present model.\\
\noindent The Eqs. (\ref{eq9}-\ref{eq11}) significantly simplify under the restriction on $\xi$ and $\eta$ given by 
\be\label{eq1a}
\eta+2\xi=0
\ee
In this case the field equations (\ref{eq9}-\ref{eq11}) contain only second derivatives of the metric and the scalar field, avoiding problems with higher order derivatives \cite{capozziello1, sushkov}. The modified Friedmann equations (\ref{eq9}) and (\ref{eq10}) take the form
\be\label{eq12}
H^2=\frac{\kappa^2}{3}\left(\frac{1}{2}\dot{\phi}^2+V(\phi)+9\xi H^2F(\phi)\dot{\phi}^2\right)
\ee
and
\be\label{eq13}
-2\dot{H}-3H^2=\kappa^2\left[\frac{1}{2}\dot{\phi}^2-V(\phi)-\xi\left(3H^2+2\dot{H}\right)F(\phi)\dot{\phi}^2-2\xi H\left(2F(\phi)\dot{\phi}\ddot{\phi}+\frac{dF}{d\phi}\dot{\phi}^3\right)\right]
\ee
where we have replaced $\eta=-2\xi$, and the equation of motion reduces to
\be\label{eq14}
\ddot{\phi}+3H\dot{\phi}+\frac{dV}{d\phi}+3\xi H^2\left(2F(\phi)\ddot{\phi}+\frac{dF}{d\phi}\dot{\phi}^2\right)
+18\xi H^3F(\phi)\dot{\phi}+12\xi H\dot{H}F(\phi)\dot{\phi}=0
\ee
Next we try to find cosmological solutions, giving rise to to accelerated expansion and acceptable behavior of the equation of state parameter (EoS). 
In the point {\bf A} of the next section we consider the general equations (\ref{eq9}) and (\ref{eq11}) in the special case when the kinetic coupling dominates over the free kinetic term. But in the rest of the paper we will study the cosmological implications of the Eqs. (\ref{eq12}) and (\ref{eq14}).\\
\noindent First note that, independently of $F(\phi)$, if we consider the asymptotic solution $\phi=\phi_0=const.$, then from Eqs. (\ref{eq9}) and (\ref{eq11}) it follows that $V=V_0=const$ and $H=H_0=\kappa\sqrt{V_0/3}$ and we get an asymptotic de Sitter behavior.\\


\section{Power-law solutions without free kinetic term}
It is of interest to derive a scalar coupling function $F(\phi)$ that
gives rise to a power-law expansion ($a\propto t^p$), as it's well known that this kind of expansion is characteristic of the evolution at the early stages of radiation and matter dominance, and also may respond for accelerated expansion.\\
Let's consider the particular case of the model (\ref{eq1}) without free kinetic term. In this case we are in the frames of above  models (\cite{sergei,allemandi,sergei1,sergei2}) which correspond to pure dark energy scalar sector. This also applies if the the slow-roll condition $\dot{\phi}^2<<V(\phi)$ is considered, which can take place in the contexts of dark energy
or inflationary cosmology. Let us consider the effects of this new kinetic coupling in the cosmological dynamics, in the case of scalar field dominance (we further neglect any background radiation or matter contribution).\\ 

\noindent{\bf A. Derivative coupling with $\eta+2\xi\neq 0$}\\

Let us then, begin with the simple case of strictly non-minimal kinetic coupling, without potential term. Introducing the new variable $\chi=\kappa^2F(\phi)\dot{\phi}^2$, and considering first the general Eqs. (\ref{eq9}-\ref{eq11}) with arbitrary $\xi$ and $\eta$ and without free kinetic and potential terms, the Friedmann equation and the equation of motion from Eqs. (\ref{eq9}) and (\ref{eq11}) take the form
\be\label{eq15}
H^2=3\xi H^2\chi+(2\xi+\eta)\dot{H}\chi-\frac{1}{2}(2\xi+\eta)H\dot{\chi}
\ee
\be\label{eq16}
\left[(4\xi+\eta)H^2+(2\xi+\eta)\dot{H}\right]\dot{\chi}
+2\left[3(4\xi+\eta)H^3+(14\xi+5\eta)H\dot{H}+(2\xi+\eta)\ddot{H}\right]\chi=0
\ee
where we have multiplied the Eq. (\ref{eq11}) by $\dot{\phi}$. Let's propose now a power law solution $H=p/t$, and replace in Eq. (\ref{eq16}) to obtain
\be\label{eq17}
\left[(4\xi+\eta)p-(2\xi+\eta)\right]\dot{\chi}+2\left[3(4\xi+\eta)p^2-(14\xi+5\eta)p+2(2\xi+\eta)\right]\chi=0
\ee
The solution to this equation is of the form
\be\label{eq18}
\chi=t^{\alpha},\,\,\,\, \alpha=\frac{3(4\xi+\eta)p^2-(14\xi+5\eta)p+2(2\xi+\eta)}{2\xi+\eta-(4\xi+\eta)p}
\ee
where we considered the condition $\chi_0=1$ at $t_0=1$. Replacing this solution in Eq. (\ref{eq15}), we obtain the following restrictions on the parameters
\be\nonumber
3(4\xi+\eta)p^2-(14\xi+5\eta)p+2(2\xi+\eta)=0,
\ee
\be\label{eq19}
p=\frac{2\xi+\eta}{3\xi-1}
\ee
giving rise to the simple restriction for accelerated expansion ($p>1$): $\xi>1/3, \xi-\eta<1$ or $\xi<1/3, \xi-\eta>1$. Replacing $p$ from the first condition (which is equivalent to $\alpha=0$) in the second one, gives an equation for $\xi$ and $\eta$, which has the three following solutions 
\be\label{eq20}
\eta=-2/3,\,\,\,\,\,\, \eta=-2\xi,\,\,\,\,\,\,\, \eta=-\xi-1
\ee
the solution $\eta=-2/3$ can not give rise to accelerated expansion, the second solution gives $p=0$, and the third solution gives accelerated expansion on the line $\eta=-\xi-1$ in the interval $0<\xi<1/3$. The scalar field can be obtained from the Eq. (\ref{eq18}) for $\alpha=0$, as follows
\be\label{eq21}
\kappa\int\sqrt{F(\phi)}d\phi=\pm t
\ee
for a given scalar function $F(\phi)$. Thus for $F(\phi)=1/\phi^2$, as considered in \cite{granda}, and taking $\alpha=0$, the scalar field is obtained as $\phi=\propto e^{\pm t/\kappa}$. The above simplifications allowed us to check that the power-law solution takes place even without the restriction (\ref{eq1a})\\
\noindent{\bf B. Derivative coupling with $\eta+2\xi=0$}\\
From now on, we will assume that the relation (\ref{eq1a}) holds. Turning to the equations (\ref{eq12})-(\ref{eq14}), the Eq. (\ref{eq12}) reduces to
\be\label{eq22}
3\xi\kappa^2F(\phi)\dot{\phi}^2=1
\ee
which can be solved in terms of the integral
\be\label{eq23}
\int\sqrt{F(\phi)}d\phi=\pm \frac{t}{\sqrt{3\xi}\kappa}
\ee
and the Eq. (\ref{eq14}) (without free kinetic term and potential) takes the simple form 
\be\label{eq24}
3H^2+2\dot{H}=0
\ee
and therefore $H=2/(3t)$, which gives a power-law corresponding to pressureless matter dominance. Note that substituting back to Lagrangian the solution (\ref{eq22}), (\ref{eq23}), we get kind of non-covariant modified gravity (for general review , see (\cite{sergei3})).\\
From Eq. (\ref{eq12}), the effective gravitational coupling can be expressed as
\be\label{eq24a}
\kappa_{eff}^2=8\pi G_{eff}= \kappa^2(1-3\xi\kappa^2\dot{\phi}^2F(\phi))^{-1}
\ee
by other hand, taking the derivative of the above expression, and using the equations for $\chi=\kappa^2\dot{\phi}^2F(\phi)$ given in (\ref{eq18}) ($\alpha=0$) and (\ref{eq22}), it follows that $\dot{G}_{eff}/G_{eff}=0$, satisfying the restrictions on the time variation of the gravitational coupling \cite{uzan1}.\\
\noindent{\bf C. Derivative coupling with potential}\\

Let's now consider the scalar field potential. We will look for the shape of the potential $V(\phi)$ corresponding to the asymptotic behavior giving rise to accelerated expansion. In this case, adopting the same variable $\chi=\kappa^2F(\phi)\dot{\phi}^2$, the Eq. (\ref{eq12}) takes the form

\be\label{eq25}
H^2=\frac{\kappa^2}{3}\frac{V(\phi)}{1-3\xi\chi}
\ee
and multiplying by $\kappa^2\dot{\phi}$ the Eq. of motion (\ref{eq14}), it's obtained
\be\label{eq26}
\kappa^2\frac{dV}{dt}+3\xi H^2\frac{d\chi}{dt}+18\xi H^3\chi+12\xi H\dot{H}\chi=0
\ee
where the first two terms of Eq. (\ref{eq14}) have been dropped due to the absence of the free kinetic term.
Taking the derivative of the potential in Eq. (\ref{eq25}), and replacing in Eq. (\ref{eq26}), we obtain
\be\label{eq27}
\xi H^2\frac{d\chi}{dt}+\xi H\dot{H}\chi-3\xi H^3\chi-H\dot{H}=0
\ee
looking for power-law solutions, we replace $H=p/t$ in Eq. (\ref{eq27}), obtaining
\be\label{eq28}
\xi t\frac{d\chi}{dt}-\xi(1+3p)\chi+1=0
\ee
considering the particular solution $\chi=\frac{1}{\xi(1+3p)}$, gives the equation for the scalar field $\phi$ 
\be\label{eq28a}
\kappa^2F(\phi)\dot{\phi}^2=\frac{1}{\xi(1+3p)}
\ee
which gives the scalar field for a given function $F(\phi)$. For instance, if we take $F(\phi)=1/\phi^2$ as in \cite{granda}, then
\be\label{eq28b}
\phi=\phi_0\exp\left(-\frac{t}{\kappa\sqrt{\xi(1+3p)}}\right)
\ee
\noindent And the time dependence of the scalar field potential is obtained from Eq. (\ref{eq25}) by replacing $H$ and $\chi$
\be\label{eq29}
V=\frac{3p^2}{\kappa^2\xi}\left(\frac{3p-2}{(3p+1)}\right)\frac{1}{t^2}
\ee
note that despite the fact that the time dependence of the potential is known, the shape of the potential depends on 
the coupling $F(\phi)$ through Eq. (\ref{eq28a}). In fact, for $F(\phi)=1/\phi^2$ as in \cite{granda},
\be\label{eq30}
V(\phi)=\frac{3p^2}{\kappa^4\xi}\left(\frac{3p-2}{(3p+1)^2}\right)\frac{1}{\log^2{\phi/\phi_0}}
\ee
\noindent Note that we have not restriction on $p$, so for $p>1$ it follows the accelerated expansion.\\
Another interesting coupling is given by $F(\phi)=e^{2\lambda\phi}/\phi_0^2$ with constant $\lambda$. Then from (\ref{eq28a}), the scalar field is obtained as follows
\be\label{eq31}
\phi=\frac{1}{\lambda}\log\left(\frac{\lambda\phi_0t}{\kappa\sqrt{\xi(1+3p)}}\right)
\ee
and the potential is given by
\be\label{eq32}
V=\frac{3\lambda^2\phi_0^2p^2(3p-2)}{\xi\kappa^4(3p+1)^2}e^{-2\lambda\phi}
\ee
this field and potential generates power-law expansion in the quintessence model (\cite{padmana,copeland}).
We can also obtain a potential of the inverse power-law type, giving rise to power-law expansion. If we consider the coupling $F(\phi)=M^{-2-n}\phi^n$, then
\be\label{eq33}
\phi=M\left(\frac{n+2}{2\kappa\sqrt{\xi(1+3p)}}\right)^{\frac{2}{n+2}}t^{\frac{2}{n+2}}
\ee
and the potential is of the form
\be\label{eq34}
V=\left(\frac{3p^2(3p-2)(n+2)^2}{4\xi\kappa^4(1+3p)^2}\right)\frac{M^{n+2}}{\phi^{n+2}}
\ee
Note that the case $n=0$ reproduces the tachyon potential giving rise to power-law expansion (see \cite{padmana,copeland})\\
\noindent The general solution of Eq. (\ref{eq28}) is given by
\be\label{eq35}
\chi=\kappa^2F(\phi)\dot{\phi}^2=t^{1+3p}+\frac{1}{\xi(1+3p)}
\ee
which after integration with respect to time, gives the expression for the scalar field as follows
\be\label{eq36}
\begin{aligned}
&\kappa\int\sqrt{F(\phi)}d\phi=\frac{2t\left(1+\xi(1+3p)t^{1+3p}\right)^{1/2}}{3(1+p)\left(\xi(1+3p)\right)^{1/2}}\\
&+\frac{t\left(\xi(1+3p)\right)^{1/2}}{3\xi(1+p)}\text{Hypergeometric2F1}\left[\frac{1}{1+3p},\frac{1}{2},1+\frac{1}{1+3p},-\xi(1+3p)t^{1+3p}\right]
\end{aligned}
\ee
The time dependence of the potential, as follows from (\ref{eq35}) and (\ref{eq25}) (with $H=p/t$) is given by 
\be\label{eq37}
V(t)=\frac{3p^2}{\kappa^2}\left(\frac{3p-2}{3p+1}t^{-2}-3\xi t^{3p-1}\right)
\ee
which gives the shape of the potential $V(\phi)$ through Eq.(\ref{eq36}). For such configuration of the scalar field and potential, and despite the fact that the potential can not be solved explicitly in terms of the field, it is possible to have power-law expansion with $p>1$. The asymptotic behavior of this potential at $t\rightarrow0$ and $t\rightarrow\infty$ is the same as discussed in \cite{granda} for $F(\phi)=1/\phi^2$. The only difference is that the shape of the potential depends on $F(\phi)$ through Eq. (\ref{eq36}).\\ 
\noindent One can use the freedom in choosing the coupling function $F(\phi)$ in order to simplify the integration in Eq. (\ref{eq35}). Making $\kappa=1$ and proposing a scalar field of the form $\phi=t$, the Eq. (\ref{eq35}) becomes
\be\label{eq38}
\sqrt{F(t)}dt=\left(t^{1+3p}+\frac{1}{\xi(1+3p)}\right)^{1/2}dt
\ee
which gives a solution
\be\label{eq38a}
F(\phi)=\left(\phi^{1+3p}+\frac{1}{\xi(1+3p)}\right)
\ee
And the corresponding potential, as seen from (\ref{eq25}) and (\ref{eq35}), can be expressed explicitly in terms of the scalar field as
\be\label{eq38b}
V(\phi)=3p^2\left(\frac{3p-2}{3p+1}\frac{1}{\phi^{2}}-3\xi \phi^{3p-1}\right)
\ee
So this coupling function gives rise to power-law expansion $a\propto t^p$, without any restriction on positive values of $p$.\\
For the particular solution Eq. (\ref{eq28a}), the restriction on the time variation of the Newtonian coupling is satisfied, as discussed at the end of the point {\bf B}. For the general solution Eq. (\ref{eq35}) we can follow the same discussion given in \cite{granda} (see Eq. (3.24) in \cite{granda}).


\section{Dynamically varying equation of state without free kinetic term}
In this case we propose an asymptotical solution for the scalar field at late times, and try to reconstruct the Hubble parameter and the potential, giving rise to dynamically changing EOS parameter, evolving according to the current observations. 
Let us write the equations (\ref{eq12}) and (\ref{eq14}) in terms of the variable $x=\log a$. The Eq. (\ref{eq12}) without free kinetic term becomes
\be\label{eq39}
H^2=\frac{\kappa^2}{3}\left(V(\phi)+9\xi H^4\phi'^2F(\phi)\right)
\ee
and the Eq. of motion (\ref{eq14}), after multiplying by $\dot{\phi}$ takes the form
\be\label{eq40}
\frac{dV}{dx}+3\xi H^4\left(2\phi'\phi''F(\phi)+\phi'^2\frac{dF}{dx} +6\phi'^2F(\phi)\right)+9\xi H^2\frac{dH^2}{dx}\phi'^2F(\phi)=0
\ee
where $\phi'=d\phi/dx$, and we used $\frac{d}{dt}=H\frac{d}{dx}$.\\
\noindent Resolving the Eq. (\ref{eq39}) with respect to $H^2$ one obtains
\be\label{eq41}
H^2=\frac{1}{6\xi\kappa^2\phi'^2F(\phi)}\left(1\pm\sqrt{1-4\xi\kappa^4 V\phi'^2F(\phi)}\right)
\ee
In this case, in order to keep $H^2$ positive, we should take care of the sing of the combination $\xi F$ (assuming that V is positive). So, if $\xi F>0$, we can take both signs of the root, and for $\xi F<0$ we take the $(-)$ root. In any case, the choice of the sign does'n influence the Eq. (\ref{eq40}), as $V$ in Eq. (\ref{eq41}) is under the square root. The simplest case of Eqs. (\ref{eq40}) and (\ref{eq41}) is obtained by considering (see \cite{granda}) 
\be\label{eq41a}
\phi'^2F(\phi)=const.=\alpha^2
\ee
here we assume that $F(\phi)>0$. Replacing in Eq. (\ref{eq41}) we get
\be\label{eq42}
H^2=\frac{\kappa^2}{3}\tilde{V}(x),\,\,\,\,\,\,  \tilde{V}(x)=\frac{1}{2\xi\alpha^2\kappa^4}\left(1\pm\sqrt{1-4\xi\alpha^2\kappa^4 V}\right)
\ee
setting $\kappa^2=1$ and replacing Eq. (\ref{eq42}) in Eq. (\ref{eq40}), we obtain the following equation for the new potential $\tilde{V}(x)$ 
\be\label{eq43}
\left(\xi\alpha^2\tilde{V}(x)-1\right)\frac{d\tilde{V}(x)}{dx}-2\xi\alpha^2\tilde{V}^2(x)=0
\ee
Integrating this equation we obtain the solution for $\tilde{V}$ as follows
\be\label{eq44}
\tilde{V}(x)=-\left(\xi\alpha^2\text{ProductLog}\left[-\frac{1}{\xi\alpha^2V_0}e^{-2x-1/(\xi\alpha^2V_0)}\right]\right)^{-1}
\ee 
using this solution in Eq. (\ref{eq42}) (with $\kappa^2=1$), and changing to the redshift variable $z$, the Hubble parameter is obtained as follows  
\be\label{eq45}   
H^2(z)=-\left(3\xi\alpha^2\text{ProductLog}\left[-\frac{1}{\xi\alpha^2V_0}e^{-1/(\xi\alpha^2V_0)}(1+z)^2\right]\right)^{-1}
\ee
Using the properties of the $\text{ProductLog}$ function, and the sign of the coupling constant $\xi$ (here $V_0>0$, and without loss of generality we assumed $F(\phi)>0$, as the kinetic coupling is given by the product $\xi F$), 
we can follow the same arguments and results given in $\cite{granda}$, which I resume here:
For positive values of the constant $\xi$ we can enter in a forbidden region for the arguments of the $\text{ProductLog}$ function. In order to avoid those regions where the  $\text{ProductLog}$ function takes complex values, we consider $\xi<0$ which will give always $H^2$ positive (see discussion in \cite{granda}).\\
\noindent From the Eq. (\ref{eq42}) and the expression for $H^2$, it follows that the effective dark energy density given by $\rho_{eff}=3H^2/\kappa^2=\tilde{V}$ tends to zero at high redshifts, validating the assumption that the dark energy was negligible at early times, when the matter contribution was relevant, but presents a Big Rip singularity at the future, revealing phantom behavior. The effective equation of state $w_{eff}=-1-\frac{1}{3}\frac{1}{H^2}\frac{dH^2}{dx}$, from Eqs. (\ref{eq42}) and (\ref{eq44}) in terms of the redshift, can be written as
\be\label{eq47}
w_{eff}=-1-\frac{2}{3\left(\text{ProductLog}\left[-\frac{1}{\xi\alpha^2V_0}e^{-1/(\xi\alpha^2V_0)}(1+z)^2\right]+1\right)}
\ee
which varies very slowly with $z$ for a given combination of $\xi, \alpha, V_0$. Fig.1 shows the behavior of the effective equation of state for two different combinations of the parameters, represented as $\gamma=\xi\alpha^2 V_0$.
\begin{center}
\includegraphics [scale=0.7]{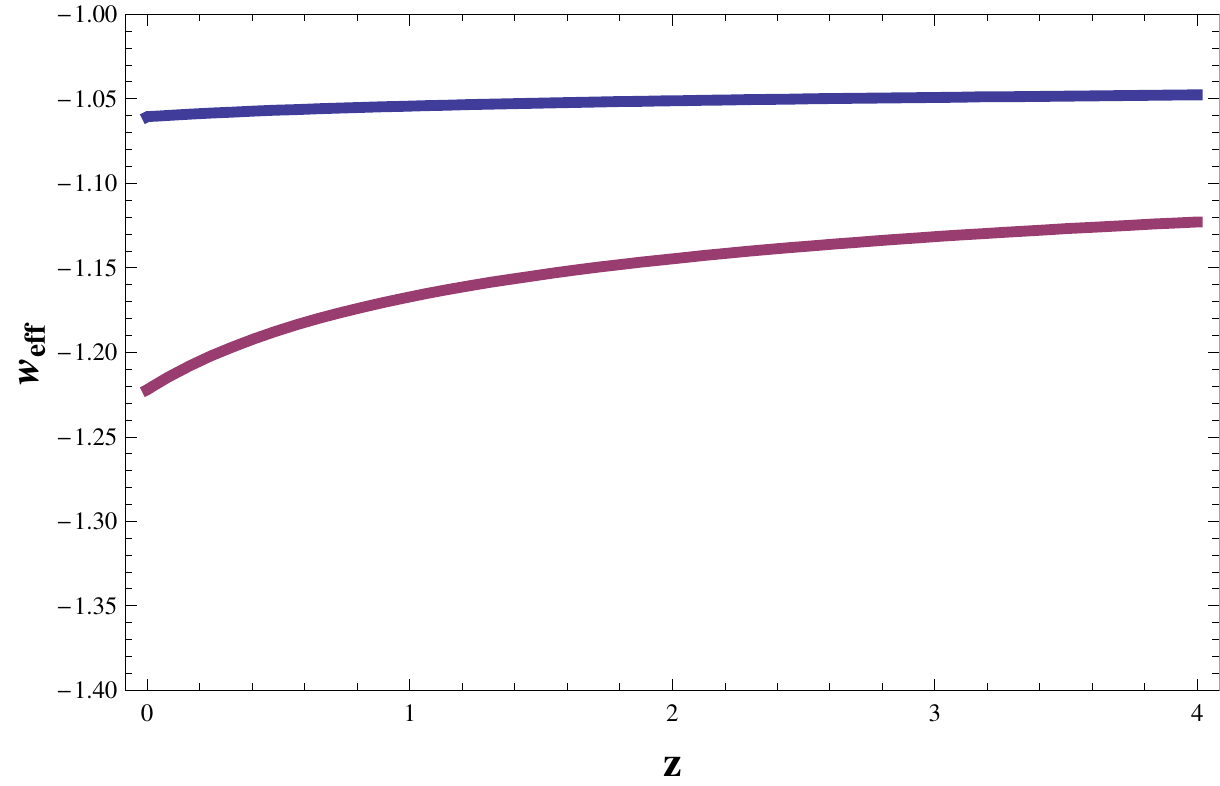}
\end{center}
\begin{center}
{Fig. 1 \it The dark energy effective equation of state parameter $w_{eff}$ versus redshift, for two combinations of the parameters $\gamma=-0.1$ (top), and $\gamma=-0.5$ (bottom), where $\gamma=\xi\alpha^2 V_0$} 
\end{center}
For the product of parameters $\gamma=\xi\alpha^2 V_0=-0.1$, the effective EoS parameter varies from $w_{eff}\approx-1.06$ at the present ($z=0$) to $-1$ at $z\rightarrow\infty$, showing very slowly variation with time, and maintaining very close to $-1$. This behavior is very close to that of the cosmological constant.\\
\noindent From Eq. (\ref{eq41a}), for a given coupling function we can find the scalar field. Thus for $F(\phi)=1/\phi^2$, solving (\ref{eq41a}), the scalar field depends on $x$ as $\phi=\phi_0 e^{-\alpha x}$. Assuming $F(\phi)=\phi^n$, the scalar field becomes $\phi=\phi_0 x^{2/(n+2)}$.\\
\noindent The time variation of the effective gravitational coupling can be found, if we use the fact that $F(\phi)\phi'^2=\alpha^2$, which allows to write 
\be\label{eq42a}
\frac{\dot{G_{eff}}}{G_{eff}}\Big|_{t_0=1}=\sigma H_0,\,\,\,\,\,\,\,\,  \sigma=\frac{6\xi\alpha^2\kappa^2\dot{H}_0}{1-3\xi\alpha^2\kappa^2H_0^2}
\ee
considering $\dot{H}_0\lesssim H_0^2$, and due to the extremely small value of $\kappa^2H_0^2=M_{Pl}^{-2}H_0^2$, is always possible to find a suitable combination of $\xi\alpha^2$ that satisfy the current constraints on $\dot{G_{eff}}/G_{eff}$ \cite{uzan1}.

\section{Cosmological solutions with free kinetic term}
In this section we include the free kinetic term and analyze some important results, specially concerning the dynamical equation of state. 

\noindent{\bf A. De Sitter Solution}\\

Multiplying the Eq. (\ref{eq14}) by $\dot{\phi}$, and replacing the product $F(\phi)\dot{\phi}^2$ from Eq. (\ref{eq12}),
the Eq. (\ref{eq14}) reduces to first order equation with respect to the variables $\psi=\dot{\phi}^2$, $H$ and $V$, and can be written as
\be\label{eq21a}
H\frac{d\psi}{dt}+\left(6H^2-\dot{H}\right)\psi+2H\frac{dV}{dt}-2\left(3H^2+\dot{H}\right)V+6\frac{H^2}{\kappa^2}\left(3H^2+2\dot{H}\right)=0
\ee
Using the fact that the variables $\psi$ and $V$ are separated, we can limit the model to the class of potentials that satisfy the restriction
\be\label{eq21b}
H\frac{dV}{dt}-\left(3H^2+\dot{H}\right)V+3\frac{H^2}{\kappa^2}\left(3H^2+2\dot{H}\right)=0
\ee
then, the equation for the field $\psi$ significantly simplifies, but still giving an interesting dynamics as can be seen bellow 
\be\label{eq21c}
H\frac{d\psi}{dt}+\left(6H^2-\dot{H}\right)\psi=0
\ee
Assuming the de Sitter solution $a(t)=a_0 e^{H_0t}$, the solution to Eqs. (\ref{eq21c}) and (\ref{eq21b}) for the scalar field and potential, are given by
\be\label{eq21d}
\phi=\phi_0\exp\left(-3H_0 t\right),\,\,\,\,\,\,\,\,  V=V_0 \frac{\phi_0}{\phi}+\frac{3H_0^2}{\kappa^2}
\ee
and replacing in Eq. (\ref{eq12}), the following expression for the coupling function is obtained
\be\label{eq21e}
F(\phi)=\frac{1}{27\xi H_0^2}\left[\frac{1}{\kappa^2\phi^2}-\frac{1}{3p^2}\log^2\left(\frac{\phi}{\phi_0}\right)\left(\frac{1}{2}+\frac{\phi_0 V_0}{9H_0^2\phi^3}+\frac{1}{3\kappa^2\phi^2}\right)\right]
\ee
In this manner, the solutions given by (\ref{eq21d}) together with the coupling (\ref{eq21e}) give an exact de Sitter (inflationary) solution, which in the case of the standard nonminimaly coupled scalar field is obtained under the known slow-roll conditions.\\
\noindent{\bf B. Dynamically varying equation of state}\\

Considering the full model with the restriction (\ref{eq1a}), we will use the freedom in choosing the coupling function in order to obtain a cosmological dynamics more close to the one expected from observations. In terms of the variable $x=\log a$ and defining the function $\theta(x)=\phi'^2$, the Eq. (\ref{eq14}) can be written as (after multiplying by $\dot{\phi}$)
\be\label{eq48}
\frac{1}{2}\frac{d}{dx}\left(H^2\theta\right)+3H^2\theta+\frac{dV}{dx}+9\xi H^2\frac{dH^2}{dx}F\theta+3\xi H^4\frac{d}{dx}(F\theta)+18\xi H^4F\theta=0
\ee
From Eq. (\ref{eq12}), changing to the variable $x$, we can write the product $F\phi'^2=F\theta$ as following 
\be\label{eq49}
F\theta=\frac{1}{3\xi\kappa^2 H^2}-\frac{\theta}{18\xi H^2}-\frac{V}{9\xi H^4}
\ee
taking the derivative of Eq. (\ref{eq49}) and replacing $F\theta$ and $d(F\theta)/dx$ into Eq. (\ref{eq48}), we arrive at the following equation involving $\theta$, $H$ and $V$
\be\label{eq50}
\begin{aligned}
&2H^4\frac{d\theta}{dx}+H^2\left(12H^2+\frac{dH^2}{dx}\right)\theta+4H^2\frac{dV}{dx}-2\left(6H^2+\frac{dH^2}{dx}\right)V\\
&+12\frac{H^2}{\kappa^2}\left(3H^2+\frac{dH^2}{dx}\right)=0
\end{aligned}
\ee
In this manner, we obtain a first order differential equation for the functions $\theta$, $H$ and $F$. In order to integrate this equation, and thanks to the fact that the functions $\theta$ and $V$ are separated, we can impose a restriction on the scalar field potential given by the equation
\be\label{eq51}
2\tilde{H}^2\frac{dV}{dx}-\left(6\tilde{H}^2+\frac{d\tilde{H}^2}{dx}\right)V+6\frac{H_0^2}{\kappa^2}\tilde{H}^2\left(3\tilde{H}^2+\frac{d\tilde{H}^2}{dx}\right)=0
\ee
which simplifies the Eq. (\ref{eq50}):
\be\label{eq52}
2\tilde{H}^2\frac{d\theta}{dx}+\left(12\tilde{H}^2+\frac{d\tilde{H}^2}{dx}\right)\theta=0
\ee
where we have defined the scaled Hubble parameter $\tilde{H}^2=H^2/H_0^2$ (note that this equivalent to divide the Eq. (\ref{eq50} by $H_0^2$). Let us propose an interesting solution for the Hubble parameter, which is important in describing the dark energy, and can fit the observations (see \cite{yin-ze})
\be\label{eq53}
\tilde{H}^2= \Omega_Ae^{-\alpha x}+\Omega_B e^{-\beta x},\,\,\,\,\, \Omega_A>0 ,\,\,\,\,\,\, \Omega_B>0
\ee
where the dimensionless parameters $\Omega_A$ and $\Omega_B$ satisfy the natural condition at the present time ($x=0$)
\be\label{eq53a}
\Omega_A+\Omega_B=1
\ee
with this proposal for $\tilde{H}^2$ we can find the functions $\theta(x)$ and $V(x)$, that satisfy the Eqs. (\ref{eq51}) and (\ref{eq52}). The constants $\Omega_A$, $\alpha$ and $\beta$ can be fixed by comparing the behavior of the effective equation of state with that expected from observational data. Replacing  $\tilde{H}^2$ in (\ref{eq52}), after integration we obtain the following expression for $\theta$
\be\label{eq54}
\theta(x)=\phi'^2=\theta_0\frac{e^{\frac{1}{2}\left(\alpha+\beta-12\right)x}}{\left(\Omega_Ae^{\beta x}+\Omega_B e^{\alpha x}\right)^{1/2}}
\ee
where $\theta_0$ is the integration constant. Replacing now $\tilde{H}^2$ in (\ref{eq51}), and rescaling the potential in the form $\tilde{V}=\kappa^2 V/H_0^2$, it is found the following expression for the scaled (dimensionless) scalar field potential
\be\label{eq55}
\begin{aligned}
&\tilde{V}(x)=C_1e^{-\frac{1}{2}(\alpha+\beta-6)x}\left(\Omega_A e^{\beta x}+\Omega_B e^{\alpha x}\right)^{1/2}+6e^{-\alpha x} \frac{\left(1+\frac{\Omega_B}{\Omega_A}e^{x(\alpha-\beta)}\right)^{1/2}}{(\alpha+6)(\alpha-2\beta-6)}\times\\
& \Big[\Omega_A(\alpha-3)(2\beta-\alpha+6)\text{Hypergeometric2F1}\left[\frac{6+\alpha}{2(\beta-\alpha)},\frac{1}{2},\frac{\alpha-2(\beta+3)}{2(\alpha-\beta)},-\frac{\Omega_B}{\Omega_A}e^{(\alpha-\beta)x}\right]+\\&
\Omega_B(\alpha+6)(\beta-3) e^{(\alpha-\beta)x}\text{Hypergeometric2F1}\left[\frac{\alpha-2(\beta+3)}{2(\alpha-\beta)},\frac{1}{2},\frac{4\beta-3\alpha+6}{2(\beta-\alpha)},-\frac{\Omega_B}{\Omega_A}e^{(\alpha-\beta)x}\right]\Big]
\end{aligned}
\ee
where the integration constant $C_1$ is now dimensionless. Integrating the square root of Eq. (\ref{eq54}), we find the scalar field as
\be\label{eq56}
\begin{aligned}
\phi=&\phi_0\left(1+\frac{\Omega_B e^{(\alpha-\beta)x}}{\Omega_A}\right)^{1/4}\frac{e^{\frac{1}{4}(\alpha+\beta-12)}}{\left(\Omega_A e^{\beta x}+\Omega_B e^{\alpha x}\right)^{1/4}}\times \\
&\text{Hypergeometric2F1}\left[\frac{\alpha-12}{4(\alpha-\beta)},\frac{1}{4},1+\frac{\alpha-12}{4(\alpha-\beta)},-\frac{\Omega_B}{\Omega_A}e^{(\alpha-\beta)x}\right]
\end{aligned}
\ee
Finally, the coupling function $F(x)$, is obtained by replacing the Eqs. (\ref{eq53}-\ref{eq55}) in (\ref{eq49}).\\
\noindent Hence, starting from the model with the action (\ref{eq1}) (with $S_m=0$), as we show bellow, one can find an exact solution that not only gives the accelerated expansion, but also gives the correct transition from decelerating phase to accelerating phase (for transition deceleration-acceleration in $f(G)$-gravity see \cite{sergeio}).\\

Expressing $\tilde{H}^2$ in terms of the redshift $z$, it takes the form
\be\label{eq59}
\tilde{H}^2=\Omega_A\left(1+z\right)^{\alpha}+\Omega_B\left(1+z\right)^{\beta}
\ee
writing the effective equation of state parameter (EoS) in the form
\be\label{eq60}
w_{eff}=-1+\frac{1}{3}\frac{(1+z)}{\tilde{H}^2}\frac{d(\tilde{H}^2)}{dz}
\ee
we can adjust the constants $\alpha$, $\beta$ and $\Omega_A$ in order to obtain the appropriate behavior, compatible with the observational data. In Fig.2 we plot the effective EoS for three different sets of the constants, and in the case corresponding to the set of parameters $\alpha=3$, $\beta=-1.49$ and $\Omega_A=0.27$ the (red) curve crosses the $-1$ divide at $z\approx 0.1$ showing quintom behavior (for a recent review on observational and theoretical status of quintom models, see \cite{yifu}). Note that all curves correctly describe the matter dominance at early times, the transition to accelerated expansion at $z_t\sim 0.5-0.7$ (see \cite{cunha}), and the current accelerated expansion with the equation of state very close to $-1$.
\begin{center}
\includegraphics [scale=1]{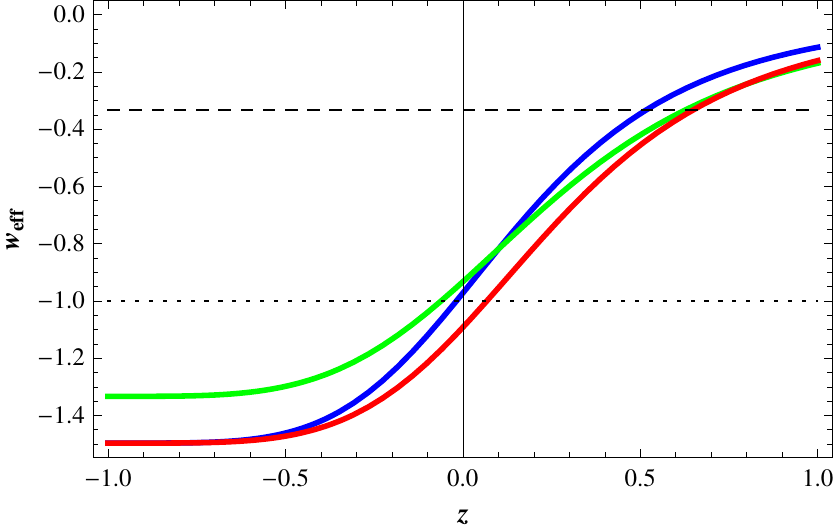}
\end{center}
\begin{center}
{Fig. 2 \it The dark energy effective equation of state $w_{eff}$ versus redshift, for three sets of parameters: green-($\alpha=3,\beta=-1,\Omega_A=0.3$), blue-($\alpha=3,\beta=-1.49,\Omega_A=0.35$) and red-($\alpha=3,\beta=-1.49,\Omega_A=0.27$)} 
\end{center}
In terms of $x$, the time variation of the gravitational coupling can be written as
\be\label{eq61}
\frac{\dot{G}_{eff}}{G_{eff}}=\frac{3\xi\kappa^2}{1-3\xi\kappa^2FH^2\theta}\frac{d}{dx}(FH^2\theta)H
\ee
replacing the numerical values of the parameters $\alpha$, $\beta$ and $\Omega_A$, in (\ref{eq53}), (\ref{eq54}) and  (\ref{eq55}), we obtain the expressions for $H$, $\theta$ and $V$ depending on $x$ and the constants $C_1$ and $C_2$ (where we replaced $\theta_0=C_2$, which is a constant of dimension $(lenght)^{-2}$). Replacing this expressions in Eq. (\ref{eq49}) and multiplying by $H^2$, we get the product $\xi\kappa^2F(x,C_1,C_2)H(x)^2\theta(x,C_2)$ , which is relevant for our analysis. After some algebra and numerical evaluation of the \text{Hypergeometric2F1} function, we arrive at the result
\be\label{eq62}
\frac{\dot{G}_{eff}}{G_{eff}}\Big|_{t_0}=\frac{3g(C_1,C_2)}{1-3f(C_1,C_2)}H_0
\ee
here we can use the constants $C_1$ and $C_2$ to guarantee that $g(C_1,C_2)\approx 10^{-2}$ or less, and $f(C_1,C_2)<<1$ in order to accomplish the restrictions imposed by the current value and the time variation of the gravitational coupling. For instance, if we take the set of parameters ($\alpha=3,\beta=-1,\Omega_A=0.3$), then $g\approx-10.71-0.34C_1+0.33\kappa^2C_2$ and $f\approx-3.42-0.11C_1-0.055\kappa^2C_2$ (note that $C_2$ appears in the dimensionless combination $\kappa^2C_2$). Instead of inequalities, we can also satisfy the restrictions by solving the linear system $g(C_1,C_2)\approx0$ and $f(C_1,C_2)\approx0$, which in this case gives $C_1\approx-30.9$ and $C_2\approx 0.2$.\\ 
The solution \ref{eq59} is very attractive, as is able to describe the current observational value of the EoS parameter, and the expected dynamical evolution of the EoS, showing not only the transition deceleration-acceleration, but also, for some values of the parameters the model shows quintom behavior. \\
In Fig.3 we have plotted the scalar field potentials corresponding to the set of parameters given in Fig.2, making the same correspondence with the colors, and with the constants $C_1$ and $C_2$ satisfying the conditions $g(C_1,C_2)\approx0$ and $f(C_1,C_2)\approx0$. Note that all the potentials present a very nice behavior, being negligible at high redshift, where we expect dark matter dominance, and becoming important as the redshift decreases, where we expect the main contribution of dark energy.  
\begin{center}
\includegraphics [scale=1]{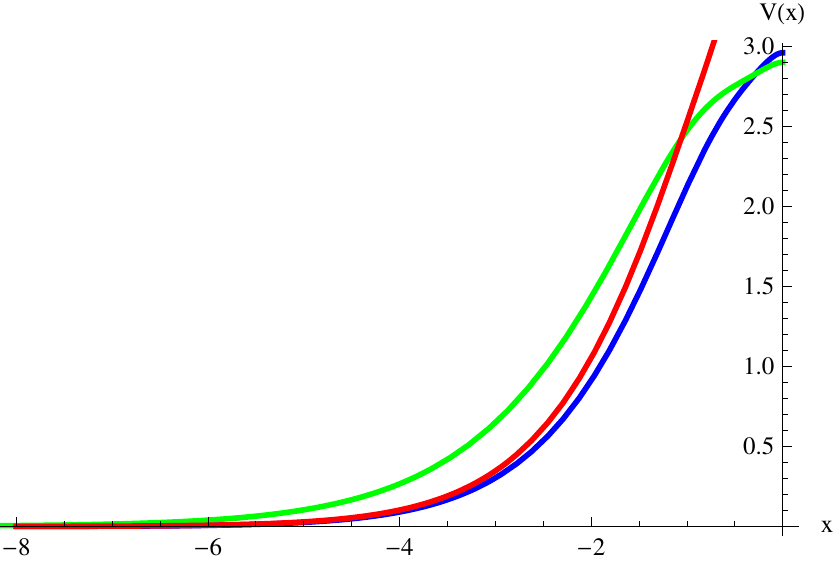}
\end{center}
{Fig. 3 \it The dark energy scalar field potentials in terms of $x=\log a$ ($a_0=1$), for the three sets of parameters:
green-($\alpha=3,\beta=-1,\Omega_A=0.3,C_1=-30.9$), blue-($\alpha=3,\beta=-1.49,\Omega_A=0.35,C_1=-1479$) and red-($\alpha=3,\beta=-1.49,\Omega_A=0.27,C_1=-1892$)} 

\section{Discussion}

We have proposed a new model of non-minimally kinetic coupled scalar field, where the kinetic  term is not only coupled to itself through the function $F(\phi)$, but to the curvature, giving rise to interesting cosmological consecuences. An interesting characteristic of the model is that the power-law behavior depends on the dynamics generated by the the nonminimal kinetic term as a whole, and the coupling function $F(\phi)$ is important in defining the time dependence of the scalar field and the shape of the potential. 
Even in the absence of a potential, the system has power-law solutions. First we considered the strictly non-minimal kinetic coupling, without the restriction on the coupling constants $\xi$ and $\eta$ (i.e. $\eta\neq-2\xi$) and have found late time power-law solutions, leading to accelerated expansion (see Eq. (\ref{eq18}-\ref{eq20})). With the restriction ($\eta+2\xi=0$), the strictly non-minimal kinetic coupling produces the effect of pressureless dark matter (\ref{eq24}), which is the behavior of the matter at early stage of the evolution. In the presence of a potential $V(\phi)$, the model presents power-law solutions giving rise to accelerated expansion, as showed in the point {\bf C} of section 3, for different couplings and potentials. In particular, the potentials for power-law solutions of minimally coupled quintessence and tachyon models are obtained in (\ref{eq32}) and (\ref{eq34}) (see \cite{padmana,copeland}). Additionally, the potential (\ref{eq37}) may be interpreted in two asymptotical cases, corresponding to early and late time behavior.\\

\noindent Looking for solutions in the presence of potential, but giving rise to dynamical effective EoS (not of the power-law type), we proposed a restriction for the scalar field and the coupling, of the form $\phi'^2F(\phi)=const.=\alpha^2$, and found the potential presented in Eq. (\ref{eq44}). According to this result, the model presents phantom behavior, but with an EoS very close to that of the cosmological constant, as seen from Fig. 1. Note from Eq. (\ref{eq42}) and Fig.1, that when the combination $\gamma=-\xi\alpha^2 V_0\rightarrow-0$, the effective EoS parameter $w_{eff}\rightarrow-1$, behaving very close to the cosmological constant. The effective EoS parameter varies from $-1$ in the far past $z\rightarrow\infty$ to $-5/3$ in the future ($z=-1$), but is worth to note that Eq. (\ref{eq42}) allows any current value of $w_{eff0}$ in the limits $-5/3<w_{eff0}<-1$, and this value is closer to $-1$ for small values of $\gamma$. For instance, for $\gamma=-0.1$, the effective EoS parameter changes from $-1.06$ to $-1$ in a practically infinite time, giving very exact description of the cosmological constant. In forthcoming work we are considering the matter contribution in order to fix the parameters of the model, according to the observational data. Note that in models with non-minimal coupling to f(R), we can consider also $f(G)$ (where $G$ is Gauss-Bonnet combination) as function multiplied to kinetic scalar term. Then it would of interest to study generalization of model (\ref{eq1}) where instead of $R$ we use function $G$ (see (\cite{sergei4})).\\

By imposing the restriction given by Eq. (\ref{eq51}) on the scalar field potential, we have found a class of solutions which describe an appropriate cosmological evolution as shown in Fig. 2, showing the transition from the deceleration to acceleration phase at $z_T\sim 0.5-0.7$, and also crossing the phantom divide at $z\sim 0.1$ (the red curve). All the presented solutions respect the restrictions imposed by the time variation of the Newtonian coupling. \\

\noindent In conclusion, the scalar field model with non-minimal kinetic coupling to itself and to the curvature, provides a new framework to describe the actual phenomenology of the dark energy. The derivative couplings provide an effective scalar density and pressure which might play an important role in the explanation of the dark energy or cosmological constant. This makes the present model very attractive for future study of the dark energy problem. 

\section*{Acknowledgments}
This work was supported by Universidad del Valle, project N° CI-7796.


\begin{thebibliography}{99}
\bibitem{riess} A.G. Riess, et al., Astron. J. 116, 1009 (1998); astron. J. 117,
707 (1999). 
\bibitem{perlmutter} S.Perlmutter \textit{et al}, Nature \textbf{391}, 51 (1998)
\bibitem{kowalski} M. Kowalski, et. al., Astrophys. Journal, \textbf{686}, p.749 (2008), arXiv:0804.4142
\bibitem{spergel1} D. N. Spergel et al., Astrophys. J. Suppl. \textbf{170}, 377 (2007); astro-ph/0603449.
\bibitem{tegmark} M. Tegmark et al. [SDSS Collaboration], Phys. Rev. \textbf{D69}, 103501 (2004) [astro-ph/0310723]
\bibitem{abazajian1} K. Abazajian et al. [SDSS Collaboration], Astron. J. 129, 1755 (2005) [astro-ph/0410239].
\bibitem{tegmark2} M. Tegmark et al. [SDSS Collaboration], Phys. Rev. \textbf{D74}, 123507 (2006), astro-ph/0608632.
\bibitem{copeland} E. J. Copeland, M. Sami and S. Tsujikawa, Int. J. Mod. Phys. \textbf{D15}
1753-1936 (2006), arXiv:hep-th/0603057
\bibitem{sahnii} V. Sahni, 	Lect. Notes Phys. \textbf{653}, 141-180 (2004), arXiv:astro-ph/0403324v3
\bibitem{padmanabhan} T. Padmanabhan, Phys. Rept \textbf{380}, 235 (2003), [hep-th/0212290].

\bibitem{green} M. B. Green, J. H. Schwarz and E. Witten, {\it 
Superstring Theory}, Cambridge University Press (1987).
\bibitem{RP}B. Ratra and P. J. E. Peebles, Phys. Rev. {\bf D37}, 3406 (1988)
\bibitem{wett} C. Wetterich, Nucl. Phys \textbf{B302}, 668 (1988).
\bibitem{shani} V. Sahni, H. Feldman and A. Stebbins, Astrophys. J. {\bf 385}, 1 (1992)
\bibitem{stein} R. R. Caldwell, R. Dave and P. J. Steinhardt, Phys. Rev. Lett. {\bf 80}, 1582
	(1998).
\bibitem{copeland97} E.J. Copeland, A.R. Liddle  and D. Wands,
 Phys. Rev. {\bf D57}, 4686 [gr-qc/9711068] (1997)	
\bibitem{stein1} I. Zlatev, L. M. Wang and P. J. Steinhardt, Phys. Rev.
Lett. 82, 896 (1999)
\bibitem{stein2} P. J. Steinhardt, L. M. Wang and
I. Zlatev, Phys. Rev. \textbf{D59}, 123504 (1999)
\bibitem{stein3} C. Armendariz-Picon, V. Mukhanov, and P. J. Steinhardt,
Phys. Rev. Lett. 85, 4438 (2000)
\bibitem{chiba} T. Chiba, T. Okabe and M. Yamaguchi, Phys. Rev. \textbf{D62}, 023511 (2000); astro-ph/9912463.
\bibitem{pad} T. Padmanabhan and T. R. Choudhury, Phys. Rev. \textbf{D66}, 081301 (2002).
\bibitem{caldwell} R. R. Caldwell, Phys. Lett. \textbf{B545}, 23-29 (2002)
\bibitem{chiba1} T. Chiba, Phys. Rev. \textbf{D60}, 083508 (1999); gr-qc/9903094
\bibitem{uzan} Jean-Philippe Uzan, Phys. Rev. \textbf{D59} (1999) 123510; gr-qc/9903004
\bibitem{wands} Damien J. Holden and David Wands, Phys. Rev. \textbf{D61}, 043506 (2000); gr-qc/9908026
\bibitem{amendola1} L. Amendola, D. Bellisai and F. Occhionero, Phys.Rev. {\textbf D47}, 4267 (1993)
\bibitem{maeda} T. Futumase et K. Maeda,
Phys. Rev. {\textbf D39}, 399 (1989).
\bibitem{turner} F.S. Accetta, D.J. Zoller, M.S. Turner, Phys. Rev. \textbf{D31}, 3046 (1985)
\bibitem{kasper} U. Kasper, Nuovo Cim. \textbf{B103}, 291 (1989)
\bibitem{easson1} D.A. Easson, JCAP, \textbf{0702}, 004 (2007); astro-ph/0608034
\bibitem{easson2} D.A. Easson, R. Gregory, Phys. Rev. \textbf{D80}, 083518 (2009); arXiv:0902.1798
\bibitem{easson3} D.A. Easson, S. Mukohyama and B.A. Powell, Phys. Rev. \textbf{D81}, 023512 (2010); arXiv:0910.1353
\bibitem{tseytlin} Hong Liu and A.A. Tseytlin, Nucl. Phys. \textbf{B533}, 88 (1998); hep-th/9804083
\bibitem{odintsov2} S. Nojiri and S.D. Odintsov, Phys. Lett. \textbf{B444}, 92 (1998); hep-th/9810008 
\bibitem{amendola2} Luca Amendola, Phys. Lett. {\bf B301}, 175 (1993); gr-qc/9302010
\bibitem{capozziello1}  S. Capozziello, G. Lambiase, Gen. Rel. Grav. {\textbf 31}, 1005 (1999); gr-qc/9901051
\bibitem{capozziello2}  S. Capozziello, G. Lambiase, H.-J.Schmidt, Annalen Phys. {\textbf 9}, 39 (2000); gr-qc/9906051
\bibitem{germani} C. Germani and A. Kehagias, arXiv:1003.2635[hep-ph]
\bibitem{caldwell1} Scott F. Daniel and Robert R. Caldwell, Class. Quant.Grav {\textbf24}, 5573 (2007); arXiv:0709.0009
\bibitem{sushkov} S.V. Sushkov, Phys. Rev. \textbf{D80}, 103505 (2009); arXiv:0910.0980
\bibitem{saridakis} E. N. Saridakis and S. V. Sushkov, Phys. Rev. \textbf{D81}, 083510 (2010); arXiv:1002.3478 
\bibitem{sergei} S Nojiri and S.D. Odintsov, Phys.Lett. {\textbf B599},137,2004, astro-ph/0403622;
PoS WC2004, 024 (2004), hep-th/0412030.
\bibitem{allemandi} G.Allemandi, A.Borowiec, M.Francaviglia and S.D. Odintsov, Phys. Rev. \textbf{D72} ,063505 (2005), gr-qc/0504057.
\bibitem{sergei1} T.Inagaki, S. Nojiri and S.D. Odintsov, JCAP, \textbf{0506} ,010 (2005), gr-qc/0504054.
\bibitem{sergei2} G.Cognola, E.Elizalde, S Nojiri and S.D. Odintsov, arXiv:0909.2747, The Open 
Astronomy Journal 2009, in press.
\bibitem{granda} L. N. Granda, arXiv:0911.3702 [hep-th]
\bibitem{sergei3} S. Nojiri and S. D. Odintsov, Int. J. Geom. Meth. Mod. Phys. {\textbf4},115, 2007; hep-th/0601213
\bibitem{uzan1} J. P. Uzan, Rev. Mod. Phys. \textbf{75}, 403 (2003)
\bibitem{padmana} T.Padmanabhan, hep-th/0204150
\bibitem{yin-ze} Yin-Zhe Ma, Nuclear Physics  \textbf{B804},  262 (2008) 
\bibitem{sergeio} S Nojiri, S D Odintsov, A. Toporensky and P.Tretyakov, arXiv:0912.2488
\bibitem{yifu} Y. Cai, E. N. Saridakis, M. R. Setare and J. Xia, arXiv:0909.2776
\bibitem{cunha} J. V. Cunha, Phys. Rev. \textbf{D79}, 047301 (2009); arXiv:0811.2379
\bibitem{sergei4} S Nojiri, S D Odintsov and P.Tretyakov, Prog. Theor. Phys. Suppl. {\bf172}, 81, 2008;
arXiv:0710.5232.
\end{thebibliography}
\end{document}